\documentclass[prl,twocolumn]{revtex4}
\newcommand{\be}{\begin{equation}}
\newcommand{\ee}{\end{equation}}
\newcommand{\bea}{\begin{eqnarray}}
\newcommand{\eea}{\end{eqnarray}}

\begin{document}

\title{Twistorial versus space--time formulations:
unification of various string models}

\author{Sergey Fedoruk,${}^1$\footnote{On leave from Ukr.
Eng. Pedag. Academy, Kharkov, Ukraine}
Jerzy Lukierski${}^2$\footnote{Supported by  KBN grant 1 P03B 01828}}
\affiliation{${}^1$Bogoliubov Laboratory of  Theoretical Physics, JINR,  141980 Dubna, Moscow Region, Russia \\
${}^2$Institute for Theoretical Physics, University of Wroc{\l}aw, pl. Maxa Borna 9, 50-204 Wroc{\l}aw, Poland}

\begin{abstract}
\noindent We introduce the $D=4$ twistorial tensionfull bosonic
string by considering the canonical twistorial 2--form in
two--twistor space. We demonstrate its equivalence to two bosonic
string models: due to Siegel (with covariant worldsheet vectorial
string momenta $P_\mu^{\,m}(\tau,\sigma)$) and the one with
tensorial string momenta $P_{[\mu\nu]}(\tau,\sigma)$. We show how to
obtain in mixed space-time--twistor formulation the
Soroka--Sorokin--Tkach--Volkov (SSTV) string model and subsequently
by harmonic gauge fixing the Bandos--Zheltukhin (BZ) model, with
constrained spinorial coordinates.

\end{abstract}

\maketitle

{\bf 1.~Introduction.} Twistors and supertwistors (see {\it
e.~g.}~\cite{PenMac,Hug,Fer}) have been recently widely
used~\cite{Shir,Bette,BC,STV,VZ} for the description of (super)
particles and (super) strings, as an alternative to space--time
approach. We stress also that recently large class of perturbative
amplitudes in $N=4$ $D=4$ supersymmetric Yang-Mills theory
\cite{12,13,14}  and conformal supergravity (see e.g. \cite{15})
were described in a simple way by using strings moving in
supertwistor space. Such a deep connection between supertwistors and
non-Abelian supersymmetric gauge fields, from other perspective
firstly observed almost thirty years ago, should promote geometric
investigations of the links between the space-time and twistor
description of the string model.
%%%!!!

In this paper we derive fourlinear twistorial classical  string action,
with target space described by two-twistor space.
%%%!!!
Our main aim is to show that the twistorial master action for
several string models
%%%!!!
which all are classical equivalent to $D=4$ Nambu-Goto string model,
can be also
%%%!!!
  described by the fundamental Liouville
2--form in two--twistor space.

Recently also there were described in $D=4$ two--twistor space
$T^{(2)}=T\otimes T$ the models describing free relativistic
massive particles with spin~\cite{Bette,FedZim,BAzLM,AFLM}. The
corresponding action was derived by suitable choice of the
variables from the following free two--twistor one--form
\be\label{1-form-x-i} \Theta^{(1)} = \Theta^{(1)}_1 +
\Theta^{(1)}_2 \ee where ($A=1,...4$, $i=1,2$; no summation over
$i$): \be\label{1-form-x} \Theta^{(1)}_i = (\bar Z^{Ai} dZ_{Ai}-
d\bar Z^{Ai} Z_{Ai}) \ee with imposed suitable constraints.

In this paper we shall study the following canonical Liouville two--form
in two--twistor space $T^{(2)}$
\be\label{2-form}
\Theta^{(2)} =  \Theta^{(1)}_1 \wedge \Theta^{(1)}_2
\ee
restricted further by suitable constraints. We shall show that from
the action which follows from~(\ref{2-form})
one can derive various formulations of $D=4$ bosonic free string theory.

We start our considerations
from the first order formulation of
the tensionfull Nambu--Goto string in flat Minkowski space which is due to
Siegel~\cite{Sieg2}~\footnote{The indices $m,n=0,1$ are vector world-sheet indices,  $h_{mn}=e_m^a e_{n a}$ is
a world--sheet metric, $e_m^a$ is the zweibein, $e_m^a e^m_b = \delta^a_b$.
The indices $a,b=0,1$ are $d=2$ flat indices.
The indices $i,j=1,2$ are $d=2$ Dirac spinor indices.  We use bar for complex conjugate quantities,
$\bar{\lambda}_{\dot\alpha}^i =(\overline{\lambda_{\alpha i}})$, and
tilde for Dirac conjugated $d=2$ spinors, $\tilde{\lambda}_{\dot\alpha}^i =
\bar{\lambda}_{\dot\alpha}^j(\rho^0)_j{}^i$.
$X^\mu(\xi)$, $\mu,\nu=0,1,2,3$, is a space--time vector and world--sheet scalar,
$P^{\,m}_\mu(\xi)$ is a space--time
vector and a world--sheet vector density.}
\begin{equation}\label{act-s}
S=\!\int\!\! d^2\xi\left[P_\mu^{\,m}\partial_m X^\mu + {\textstyle
\frac{1}{2T}}(-h)^{-1/2}h_{mn}P_\mu^{\,m} P^{\,\mu\,n}\right]\,.
\end{equation}
The kinetic part of the action~(\ref{act-s})
is described equivalently by the two--form
\begin{equation}\label{Th-2}
\tilde\Theta^{(2)} = P_\mu \wedge d X^\mu
\end{equation}
where
$P_\mu =P_\mu^{\,m}\epsilon_{mn} d \xi^n$, $d X^\mu = d \xi^m \partial_m X^\mu$
{\it i.~e.} in Siegel formulation the pair $(P_\mu^{\,0},P_\mu^{\,1})$
of generalized string momenta
are represented by a one--form.

If we apply to~(\ref{act-s}) the string generalization of the Cartan--Penrose
formula on curved world sheet [25]
\begin{equation}\label{P-res-st}
P_{\alpha\dot\alpha}^{\,m}=e\,
\tilde{\lambda}_{\dot\alpha}\rho^m\!\lambda_{\alpha} =e
e^m_a\tilde{\lambda}_{\dot\alpha}^i(\rho^a)_i{}^j\lambda_{\alpha j}\,.
\end{equation}
we shall obtain the SSTV bosonic string model~\cite{SSTV}
\begin{equation}\label{SSTV}
S=\!\int\!\! d^2\xi \, e\left[\tilde{\lambda}_{\dot\alpha}
\rho^m\!\lambda_{\alpha
}\, \partial_m X^{\dot\alpha\alpha}+ {\textstyle \frac{1}{2T}}\,(\lambda^{\alpha i}
\lambda_{\alpha i}) (\tilde{\lambda}_{\dot\alpha}^j\tilde{\lambda}^{\dot\alpha}_j)\right]
\end{equation}
where $\sqrt{-h}=e=\det(e_m^a)=-\frac{1}{2}\epsilon^{mn}\epsilon_{ab}e_m^a e_n^b$.
Further we shall discuss the local gauge freedom in the spinorial sector
of~(\ref{SSTV}) and consider the suitable gauge fixing.
We shall show that by suitable constraints in spinorial space
we obtain the BZ formulation~\cite{BZ} which interprets
the $D=4$ spinors $\lambda_{\alpha i}$, $\bar{\lambda}_{\dot\alpha}^i$ as the
spinorial Lorentz harmonics. Finally we shall derive
the second order action for
twistorial string model described
by the two--form~(\ref{2-form}).

Further we shall consider the bosonic string model with
tensorial momenta
obtained from the Liouville two--form~\cite{Gur,GZ}
\begin{equation}\label{Th-2-2}
\tilde{\tilde\Theta}{}^{(2)} = P_{\mu\nu} d X^\mu \wedge d X^\nu\,.
\end{equation}
Such a model is directly related with the interpretation of strings as dynamical
world sheets with the surface elements
\begin{equation}\label{els}
d S^{\mu\nu} = d X^\mu \wedge d X^\nu =
\partial_m X^\mu \partial_n X^\nu \epsilon^{mn} d^2\xi \,.
\end{equation}
If we introduce the composite formula for
$P_{\alpha\beta}=P_{\mu\nu}\sigma^{\mu\nu}_{\alpha\beta}$, $\bar P_{\dot\alpha\dot\beta}=-P_{\mu\nu}
\sigma^{\mu\nu}_{\dot\alpha\dot\beta}$ in terms
of spinors (see also~\cite{GZ})
by passing to
the first order action we obtain the mixed spinor--space-time
SSTV and BZ string formulations.
We see therefore that both bosonic string models,
based on~(\ref{Th-2-2}) and (\ref{Th-2}), lead via SSTV to the
purely twistorial bosonic string with the null twistor constraints and the
constraint determining the string tension $T$
\begin{equation}\label{P2-tens}
P^{\mu\nu}P_{\mu\nu} =-{\textstyle
\frac{T^2}{4}}  \quad \leftrightarrow \quad
|\lambda_{\alpha 1} \lambda^\alpha_2 |^2 ={\textstyle
\frac{T^2}{4}}\,.
\end{equation}

If we wish to obtain the BZ formulation one should introduce in place
of~(\ref{P2-tens}) two constraints
\begin{equation}\label{P2-tens-2}
\lambda_{\alpha 1} \lambda^\alpha_2 ={\textstyle
\frac{T}{2}}\,, \qquad
\bar\lambda_{\dot\alpha}^1 \bar\lambda^{\dot\alpha 2}  ={\textstyle
\frac{T}{2}}
\end{equation}
providing the particular solution of the constraint~(\ref{P2-tens}).

\bigskip
{\bf 2.~Siegel bosonic string.}
Equations of motion following from the action~(\ref{act-s}) are
\begin{eqnarray}
\partial_m P^{\,m}_\mu=0\,, \label{1}\\
P^{\,m}_\mu=
-T(-h)^{1/2}h^{mn}\partial_n X_\mu\,, \label{2}\\
P^{\,m}_\mu P^{\,n
\mu}-{\textstyle
\frac{1}{2}}h^{mn}h_{kl}P^{\,k}_\mu P^{\,l
\mu}=0 \,. \label{3}
\end{eqnarray}
If we solve half of the equations of motion~(\ref{2}) without time derivatives
\begin{equation}\label{p1}
P_\mu^1=-\rho P_\mu-\lambda\, T X_\mu^\prime
\end{equation}
where $P_\mu^0=P_\mu$ denotes the
string momentum and
$\lambda=\frac{\sqrt{-h}}{h_{11}}$, $\rho=\frac{h_{01}}{h_{11}}$,
the action~(\ref{act-s}) takes the form
\begin{equation}\label{action-1}
S\!=\!\!\int\!\! d^2\xi\left[\!P_\mu\dot{X}^\mu-\lambda{\textstyle
\frac{1}{2}}(T^{-1}\!P_\mu^2+TX_\mu^{\prime\, 2})-\rho P_\mu \!X^{\prime \mu}
\!\right].
\end{equation}
It is easy to see that~(\ref{action-1}) describes the phase space formulation
of the tensionfull Nambu--Goto string
\begin{equation}\label{action-NG}
S=-T\int d^2\xi \sqrt{-g^{(2)}}
\end{equation}
where $g^{(2)}$ is the determinant of the induced  $D=2$ metric
\begin{equation}\label{ind-metr}
g_{mn}= \partial_m X^\mu\partial_n X_\mu\,,
\end{equation}
$T$ is the string tension, and the string Hamiltonian (see~(\ref{action-1}))
is described by a summ
of first class constraints generating Virasoro algebra.

By substitution of equations of
motion~(\ref{2}) into the Siegel action~(\ref{act-s})
one obtains the Polyakov action
\begin{equation}\label{action-2}
S=-{\textstyle \frac{T}{2}}\int d^2\xi
(-h)^{1/2}h^{mn}\partial_m X^\mu\partial_n X_\mu\,.
\end{equation}

Note that the equations~(\ref{3}) describe the Virasoro first class
constraints.

\bigskip
{\bf 3.~SSTV string model and its restriction to BZ model.}
In order to obtain from the action~(\ref{act-s}) the mixed
spinor--space-time action~(\ref{SSTV}) we should eliminate the
fourmomenta $P_\mu^m$ by means of the formula~(\ref{P-res-st}).
We obtain that
the second term in string action~(\ref{act-s}) takes the
form
\begin{equation}\label{2-term}
{\textstyle \frac{1}{2T}}(-h)^{-1/2}h_{mn}P_\mu^{\,m} P^{\,n
\mu}={\textstyle \frac{1}{2T}}\,e\,
(\lambda^{\alpha i}\lambda_{\alpha i})
(\tilde{\lambda}_{\dot\alpha}^j\tilde{\lambda}^{\dot\alpha}_j)
\end{equation}
where we used ${\rm Tr}(\rho^m \rho^n) =2h^{mn}$. Note that
$\tilde{\lambda}_{\dot\alpha}^i \tilde{\lambda}^{\dot\alpha}_i =
\bar{\lambda}_{\dot\alpha}^i \bar{\lambda}^{\dot\alpha}_i$.

Putting~(\ref{P-res-st}) and (\ref{2-term}) in the action~(\ref{act-s}) we
obtain the SSTV string action~(\ref{SSTV}) which
provides the mixed space-time--twistor formulation of bosonic string.
We stress that in SSTV formulation the twistor spinors $\lambda_{\alpha i}$
are not constrained. Further, the algebraic field equation~(\ref{3}) after
substitution~(\ref{P-res-st}) is satisfied as an identity.

Calculating from the action~(\ref{SSTV}) the momenta
$\pi^{\alpha i}$, $\bar\pi^{\dot\alpha}_i$, $p^{(e)}{}^m_a$ conjugate to the variables
$\lambda_{\alpha i}$, $\bar\lambda_{\dot\alpha}^i$, $e^a_m$ one can introduce the following two first class constraints
\begin{eqnarray}\label{F-g}
F&=&\lambda_{\alpha i}\pi^{\alpha i} +\bar\lambda_{\dot\alpha}^i
\bar\pi^{\dot\alpha}_i - 2 e^a_m p^{(e)}{}^m_a \approx 0 \,,
\\
G&=&i(\lambda_{\alpha i}\pi^{\alpha i} -\bar\lambda_{\dot\alpha}^i
\bar\pi^{\dot\alpha}_i) \approx 0 \label{G-g}
\end{eqnarray}
generating the following local transformations:
$$
\lambda^\prime_{\alpha i} =e^{i(b+ic)}\lambda_{\alpha i},\quad
\bar\lambda_{\dot\alpha}^{\prime\, i} =e^{-i(b-ic)}\bar\lambda_{\dot\alpha}^i
,\quad
e^{\prime\, a}_m =e^{2c}e^{a}_m\,.
$$
In particular one can fix the real parameters $b$, $c$ in such a way that we obtain
the constraints~(\ref{P2-tens-2}).
The relations~(\ref{P2-tens-2})
can be rewritten in $SU(2)$--covariant way as follows
%%%!!!
 (we recall that $T$ is real)

\begin{equation}\label{nor2}
A = \lambda^{\alpha i} \lambda_{\alpha i}-
T = 0 \,, \qquad
\bar{A} =
\bar\lambda_{\dot\alpha}^{i} \bar\lambda^{\dot\alpha}_i - T  = 0 \,.
\end{equation}
%%%!!!
If we introduce the variables $v_{\alpha i} ={\textstyle\sqrt{\frac{2}{T}}}\, \lambda_{\alpha i}$,
$\bar v_{\dot\alpha}^i = {\textstyle\sqrt{\frac{2}{T}}}\, \bar\lambda_{\dot\alpha}^i$
we get the orthonormality relations for the spinorial
Lorentz harmonics~\cite{BZ}.

If we impose the constraints~(\ref{P2-tens-2}) the model~(\ref{SSTV})
can be rewritten in the following way
%\begin{widetext}
\begin{equation}\label{act-harm}
S=\int d^2\xi \left[e e^m_a \tilde{\lambda}_{\dot\alpha}^i
(\rho^a)_i{}^j\lambda_{\alpha
j}\, \partial_m X^{\dot\alpha\alpha}+ {\textstyle \frac{T}{2}}\,e
+ \Lambda A + \bar{\Lambda} \bar{A}
\right]
\end{equation}
%\end{widetext}
where the spinors $\lambda$, $\bar\lambda$ are constrained by the
 relations~(\ref{nor2}), which are
 imposed additionally in~(\ref{act-harm}) by the Lagrange multipliers.
It is easy to see that introducing the light cone notations
on the world sheet for the zweibein
$e_m^{++} = e_m^{0}+e_m^{1}$, $e_m^{--} = e_m^{0}-e_m^{1}$
and following  Weyl representation for Dirac matrices in two dimensions
we obtain string action in the form used by Bandos and Zheltukhin (BZ model)~\cite{BZ,BAM,Uv}.

\bigskip
{\bf 4.~Purely twistorial formulation.}
Let us introduce second half of twistor coordinates $\mu^{\dot\alpha}_i$,
$\bar{\mu}^{\alpha i}$ by employing Penrose incidence relations
generalized for string
\begin{equation}\label{Pen-inc}
\mu^{\dot\alpha}_i=X^{\dot\alpha\alpha}\lambda_{\alpha i} \,,\qquad
\bar{\mu}^{\alpha i}=\bar{\lambda}_{\dot\alpha}^i
X^{\dot\alpha\alpha}\,.
\end{equation}

Incidence relations~(\ref{Pen-inc}) with real space--time string
position $X^{\dot\alpha\alpha}$ imply
that the twistor variables satisfy the constraints
\begin{equation}\label{V}
V_i{}^j \equiv \lambda_{\alpha
i}\bar{\mu}^{\alpha j} - \mu^{\dot\alpha}_i
\bar{\lambda}_{\dot\alpha}^j \approx 0
\end{equation}
which are antiHermitian
($(\overline{V_i{}^j})=-V_j{}^i$).

Let us insert the relations~(\ref{Pen-inc}) into (\ref{act-harm}).
Using
$$
P_{\alpha\dot\alpha}^{\,m}\partial_m X^{\dot\alpha\alpha}=
{\textstyle \frac{1}{2}}e e_a^m \left[\tilde{\lambda}_{\dot\alpha}\rho^a\partial_m
\mu^{\dot\alpha}
-\tilde{\mu}^\alpha\rho^a\partial_m\lambda_\alpha
\right] + {\it c.c.}
$$
we obtain the first order string action in
twistor formulation
%\begin{widetext}
\begin{eqnarray}
S\!\! & = & \!\! \int\!\! d^2\xi
 %%%&
  \!\!\left\{  {\textstyle \frac{1}{2}}
e e_a^m \!\left[\tilde{\lambda}_{\dot\alpha}\rho^a\partial_m
\mu^{\dot\alpha}
-\tilde{\mu}^\alpha\rho^a\partial_m\lambda_\alpha
+ {\it c.c.}\right] \right.\!\!    \nonumber\\
%  \left.
%%%%%%%%%%%%%%%%
& &
+\, {\textstyle \frac{T}{2}}\,e
 + \Lambda A + \bar{\Lambda} \bar{A}
 + \Lambda_j{}^i (\lambda_{\alpha
i}\bar{\mu}^{\alpha j} - \mu^{\dot\alpha}_i
\bar{\lambda}_{\dot\alpha}^j)
%\right\}
\Big\}
\label{action-tw}
\end{eqnarray}
%\end{widetext}
where $\Lambda, \bar{\Lambda}$,  $\Lambda_j{}^i$ are the Lagrange multipliers
($(\overline{\Lambda_j{}^i})=-\Lambda_i{}^j$).
%%%%!!!and
%%%!!! the spinors $\lambda$, $\bar \lambda$ are restricted by the constraints~(\ref{nor2}).

The variation with respect to zweibein $e_m^a$
of the action~(\ref{action-tw})
gives the equations (we use that $ee^m_a=-\epsilon_{ab}\epsilon^{mn} e_n^b$)
\begin{equation} \label{resol}
e_m^a =-{\textstyle\frac{1}{T}}
\left(\tilde{\lambda}_{\dot\alpha}\rho^a\partial_m
\mu^{\dot\alpha}
-\tilde{\mu}^\alpha\rho^a\partial_m\lambda_\alpha
\right) + {\it c.c.}\,.
\end{equation}
For compact notation we introduce the string twistors
$$
Z_{Ai}=(\lambda_{\alpha i}, \mu_i^{\dot\alpha}), \,\,\,\,
\bar Z^{Ai}=(\bar\mu^{\alpha i}, -\bar\lambda_{\dot\alpha}^{i} ), \,\,\,\,
\tilde Z^{Ai}=\bar Z^{Aj}(\rho^0)_j{}^i.
$$
Then
\begin{equation}\label{e-am}
e_m^a =-{\textstyle\frac{1}{T}}
\left[\partial_m\tilde Z^{A i}
(\rho^a)_i^{\ \ j}  Z_{A j } - \tilde Z^{A i }
(\rho^a)_i^{ \ \ j} \partial_m Z_{A j}
\right]
\end{equation}
and the constraints~(\ref{V}) can be rewritten as
\begin{equation}\label{V-Z}
V_i{}^j = Z_{Ai}\bar Z^{Aj}\approx 0\,.
\end{equation}

Substituting~(\ref{e-am}) and~(\ref{V-Z})
in the action~(\ref{action-tw}) we obtain our
basic twistorial string action:
\begin{widetext}
\begin{eqnarray}
S \!  = \!\int\! d^2\xi  \!\Bigg\{ {\textstyle \frac{1}{4T}}
\epsilon^{mn}\epsilon_{ab}
\left[\partial_m
\tilde Z^{A i}
(\rho^a)_i ^{  \ j} Z_{A j}
- \tilde Z^{A i} (\rho^a)_i ^{ \ j} \partial_m
Z_{A j}
\right]\left[\partial_n\tilde Z^{B i}
 (\rho^b)_i^{ \ j} Z_{B j}
  - \tilde Z^{B i}
  (\rho^b)_i^{ \ j} \partial_n Z_{B j}
\right]
\nonumber \\
+
\Lambda A + \bar{\Lambda} \bar{A} +
\Lambda_j{}^i
%(Z_{A i}\bar Z^{A j})
V_i^{\  j}
\Bigg\}
\,. \label{action-tw-2}
\end{eqnarray}
\end{widetext}

Using explicit form of $D=2$ Dirac matrices
we can see that the first term in the
action~(\ref{action-tw-2}) equals to
$$
 {\textstyle \frac{1}{T}}\epsilon^{mn}
\!\!\left[\partial_m\bar Z^{A1}\! Z_{A1} \!-\! \bar Z^{A1} \partial_m Z_{A1}
\right]\!
\left[\partial_n
 \bar Z^{B2}\! Z_{B2} \!-\! \bar Z^{B2} \partial_n Z_{B2}\right]
$$
i.e. the action~(\ref{action-tw-2}) is induced on the world--sheet by the canonical
2--form~(\ref{2-form}) with supplemented constraints~(\ref{nor2}) and~(\ref{V}).

\bigskip
{\bf 5.~From SSTV action to tensorial momentum formulation.}
The zweibein $e_m^a$ can be expressed from the
action~(\ref{SSTV}) as follows ($(\lambda\lambda) \equiv \lambda^{\alpha i}
\lambda_{\alpha i}$,
$(\bar\lambda\bar\lambda) \equiv \bar{\lambda}_{\dot\alpha}^i \bar{\lambda}^{\dot\alpha}_i$)
\begin{equation} \label{resol-tens}
e_m^a ={\textstyle\frac{2T}{(\lambda\lambda)(\bar\lambda\bar\lambda)}}\,\,
\,\tilde{\lambda}_{\dot\alpha}^i
(\rho^a)_i{}^j\lambda_{\alpha
j}\,\, \partial_m X^{\dot\alpha\alpha}
\end{equation}

Substitution of the relation~(\ref{resol-tens}) in the action~(\ref{action-tw})
provides the following string action
\begin{widetext}
\begin{equation}\label{act-tw-ten1}
S=T\int d^2\xi\,[(\lambda\lambda)(\bar\lambda\bar\lambda)]^{-1}
\epsilon_{ab}(\tilde{\lambda}_{\dot\alpha}
\rho^a\lambda_{\alpha})\,
(\tilde{\lambda}_{\dot\beta}
\rho^b\lambda_{\beta})\,\epsilon^{mn}
\partial_m X^{\dot\alpha\alpha} \partial_n X^{\dot\beta\beta}
\end{equation}
\end{widetext}

Using identities for $D=2$ Dirac matrices and the relation
$$
\epsilon^{mn}
\partial_m X^{\dot\alpha}_{\alpha} \partial_n X^{\dot\beta}_{\beta}=
\epsilon^{mn}
\partial_m X^{[\dot\alpha}_{(\alpha} \partial_n X^{\dot\beta]}_{\beta)}+
\epsilon^{mn}
\partial_m X^{(\dot\alpha}_{[\alpha} \partial_n X^{\dot\beta)}_{\beta]}
$$
after contractions of spinorial
indices we obtain the action
\begin{equation}\label{tens-3}
S\!=\!\!{\textstyle\sqrt{2}}\!\!\int \!\!d^2\xi\epsilon^{mn}\!\!
\left(\! P_{\alpha\beta}\,
\partial_m X^{\dot\gamma\alpha} \partial_n X_{\dot\gamma}^{\beta}\!+\!
\bar P_{\dot\alpha\dot\beta}
\,\partial_m X^{\dot\alpha\gamma} \partial_n X^{\dot\beta}_{\gamma}\!
\right)
\end{equation}
where the composite second rank spinors
\begin{equation}\label{tens-mom}
P_{\alpha\beta}={\textstyle\frac{\sqrt{2}T}{({\lambda}{\lambda})}}\,
\lambda_{(\alpha}^1\lambda_{\beta)}^2\,, \qquad
\bar P_{\dot\alpha\dot\beta}={\textstyle\frac{\sqrt{2}T}{(\bar{\lambda}\bar{\lambda})}}\,
\bar{\lambda}_{(\dot\alpha}^1 \bar{\lambda}_{\dot\beta)}^2\,.
\end{equation}
satisfy the constraints
\begin{equation}\label{tens-mom-con}
P^{\alpha\beta}P_{\alpha\beta}=- {\textstyle\frac{T^2}{4}}\,, \qquad
\bar P^{\dot\alpha\dot\beta}\bar P_{\dot\alpha\dot\beta}=-{\textstyle\frac{T^2}{4}}\,.
\end{equation}
Using fourvector notation
%(see~(\ref{s-as}))
the relations~(\ref{tens-mom-con}) take the form
\begin{equation}\label{tens-mom-con1}
P^{\mu\nu}P_{\mu\nu} = -{\textstyle\frac{T^2}{4}}\,, \qquad
P^{\mu\nu}\tilde P_{\mu\nu} =0
\end{equation}
where $\tilde P_{\mu\nu}=\frac{1}{2}\epsilon_{\mu\nu\lambda\rho}P^{\lambda\rho}$.

The action~(\ref{tens-3}) is the Ferber--Shirafuji form of the string action
with tensorial momenta
\begin{equation}\label{tens-4}
S\!=\!\!{\textstyle\sqrt{2}}\!\!\int\!\! d^2\xi\!\!
\left[P_{\mu\nu}\,
\partial_m\! X^{\mu} \partial_n\! X^{\nu}\!\epsilon^{mn} \!-\!
\Lambda(P^{\mu\nu}\!P_{\mu\nu} \!+\! {\textstyle\frac{T^2}{4}})\!
\right]\!.
\end{equation}
Expressing $P_{\mu\nu}$ by its equation of motion, we get
\begin{equation}\label{p-mn}
P^{\mu\nu}={\textstyle\frac{1}{2\Lambda}}\Pi^{\mu\nu}\,,  \qquad
\Pi^{\mu\nu}\equiv\epsilon^{mn}\partial_m X^{\mu} \partial_n X^{\nu}\,.
\end{equation}
After substituting~(\ref{p-mn}) in the action~(\ref{tens-4}) we obtain the second--order action (see {\it e.g.}~\cite{BDM})
\begin{equation}\label{tens-5}
S={\textstyle\frac{1}{2\sqrt{2}}}\int d^2\xi\,
\left[ \Lambda^{-1}
\Pi^{\mu\nu}\Pi_{\mu\nu} -
\Lambda T^2
\right]\,.
\end{equation}
Eliminating further $\Lambda$ and using that (see also~(\ref{ind-metr}))
\be
\Pi^{\mu\nu}\Pi_{\mu\nu} =2 \det (g_{mn})
\ee
we obtain the Nambu--Goto string action~(\ref{action-NG}).

It is important to notice
 that the solution~(\ref{p-mn})  satisfies the constraint $P^{\mu\nu}\tilde P_{\mu\nu}=0$ as an identity.
We see therefore that in the action~(\ref{tens-4}) it
is sufficient to impose by the Lagrange multiplier
only the first constraint~(\ref{tens-mom-con1}).

\bigskip
{\bf 6.~Conclusions.}
We have shown the equivalence of five formulations of $D=4$ tensionfull bosonic string:\\
\noindent $\bullet$ two space-time formulations, with vectorial string momenta (see~(\ref{act-s}))
 and tensorial ones (see~(\ref{tens-4}));\\
\noindent $\bullet$ two mixed twistor--space-time SSTV (see~(\ref{SSTV}))
and BZ (see~(\ref{act-harm})) models; \\
\noindent $\bullet$ the generic pure twistorial formulation
with the action given
by the formula~(\ref{action-tw-2}).

Following the massive relativistic particle case
(see~\cite{FedZim,BAzLM,AFLM}) the main tools in the equivalence
proof are the string generalizations of
Cartan--Penrose string momenta (see~(\ref{P-res-st})
and~(\ref{tens-mom}))
and the incidence relations~(\ref{Pen-inc}). The
action~(\ref{action-tw}) in conformal gauge $e_m^a =\delta_m^a$
is the commonly used bilinear action for twistorial string.

%%%!!!
We would like to stress that the model (31) is substantially different
from the one proposed by Witten et al [9--11]. In
Witten twistor string model
 described by $CP(3|4)$ ( $N=4$ supertwistor) $\sigma$-model
the targed space is described by a single supertwistor,
and the Penrose incidence relation, introducing space-time coordinates appears only
after \underline{quantization}, as the step permitting the space-time
interpretation of holomorphic twistorial fields.
In our approach composite space-time variables enter already into the formulation of
 \underline{classical}
string model, in a way enforcing the complete equivalence of classical twistorial string
and Nambu-Goto action provided that we treat the space-time target coordinates as
2-twistor composites.

In this paper we restricted the presentation to the case of $D=4$ bosonic
string. The generalization to  $D=6$ is rather straightforward; the extension
 to  $D=10$ requires clarification how to introduce the  $D=10$ conformal spinors,
  i.e.  $D=10$ twistors. Other possible generalizations are the following:
%%%!!!

\noindent {\it i}\,)
If we quantize canonically the model~(\ref{action-tw})
one can show that the PB of the constraints $V_i{}^j$
satisfy the internal $U(2)$ algebra (see~\cite{Siegel}).
 One can introduce, contrary to
 (\ref{V}), nonvanishing $V_i{}^j$.
The degrees of freedom described by
$V_i{}^j$ can be interpreted (see also~\cite{Hug,BAzLM,AFLM})
as introducing on the string the local density of covariantly described
spin components and electric charge;\\
\noindent {\it ii}\,) We presented here the links between various bosonic string models.
Introducing two--supertwistor space and
following known supersymmetrization techniques
(see~\cite{BAM,Uv}) one can extend the presented equivalence
proofs to the relations between different superstring formulations
 with manifest world-sheet supersymmetry which involved the twistor variables
  (see e.g. \cite{Berkovits}--\cite{Galperin}).

%%%!!!
\noindent {\it  iii} ) Particularly interesting would be the twistorial formulation of
   $D=4$  $N=4$ Green-Schwarz superstring, which should be derivable by
   dimensional reduction from  $D=10$,  $N=1$ Green-Schwarz superstring. Such
   twistorial  $D=4$,  $N=4$ supersting model could be in our formulation the
   counterpart of twistorial  $N=4$ superstring considered in [9--11].
%%%!!!

\bigskip
{\bf Acknowlegments.}
We would like to
thank E.Ivanov and D. Sorokin  for valuable remarks.
S.F. wishes to thank Institute for Theoretical Physics, Wroc{\l}aw University and
Institut f{\" u}r Theoretische Physik, Universit{\" a}t Hannover, for kind hospitality
 and for financial support.
The work of S.F. was partially supported by the RFBR grant 06-02-16684 and
the grants from Bogoliubov--Infeld and Heisenberg-Landau programs.

\end{document}